%% file: main.tex
\documentclass[10pt,conference]{IEEEtran}
\AtBeginDocument{%
  \providecommand\BibTeX{{%
    Bib\TeX}}}

\usepackage{graphicx} %
\usepackage{multirow}
\usepackage{algorithm}
\usepackage{algpseudocode}
\usepackage{booktabs}
\usepackage{tcolorbox}
\usepackage{enumitem}
\usepackage{arydshln}
\usepackage{xspace}
\usepackage{cite}
\usepackage{amsmath,amssymb,amsfonts}
\usepackage{textcomp}
\usepackage{xcolor}
\usepackage{url} 
\usepackage{subcaption}
\usepackage{balance}
\def\BibTeX{{\rm B\kern-.05em{\sc i\kern-.025em b}\kern-.08em
    T\kern-.1667em\lower.7ex\hbox{E}\kern-.125emX}}
\newcommand{\boxmargin}{1mm}

\definecolor{lightcyan}{rgb}{0.88, 1.0, 1.0}
\colorlet{mythmback}{lightcyan!40!white}
\newtcolorbox{boxEnv}{
colback=mythmback,coltitle=blue,colframe=mythmback,
center,
width=\linewidth,
boxrule=0.5pt,
left=5pt,right=0pt,
top=2pt,bottom=2pt,
before skip=10pt, after skip=10pt
}

\newtcolorbox{mybox}{
    colback=gray!15!white,
    arc = 0pt, outer arc = 0pt,
    boxsep=0pt, left = 3pt, right = 0pt, top = 0pt, bottom = 0pt, 
    leftrule=3pt, bottomrule=0pt,toprule=0pt, rightrule=0pt,
    left = \boxmargin, right = \boxmargin, top = \boxmargin, bottom = \boxmargin
}
\newcommand{\appname}{{\textsc{FGit}}\xspace}
\newcommand{\ess}{\textit{error-sensitive segments}\xspace}

\title{\appname: \underline{F}ault-\underline{G}uided F\underline{i}ne-\underline{T}uning for Code Generation}

\author{
    \IEEEauthorblockN{
        Lishui Fan\IEEEauthorrefmark{2},
        Zhongxin Liu\IEEEauthorrefmark{2}\IEEEauthorrefmark{1},
        Haoye Wang\IEEEauthorrefmark{3},
        Lingfeng Bao\IEEEauthorrefmark{2},
        Xin Xia\IEEEauthorrefmark{2},
        Shanping Li\IEEEauthorrefmark{2}
    }
    \IEEEauthorblockA{
        \IEEEauthorrefmark{2}\textit{The State Key Laboratory of Blockchain and Data Security, Zhejiang University, China} \\
        \IEEEauthorrefmark{3}\textit{Hangzhou City University, China} \\
        \{flscode, liu\_zx, lingfengbao, shan\}@zju.edu.cn, wanghaoye@hzcu.edu.cn, xin.xia@acm.org
    }
}

\date{February 2025}

\begin{document}

\maketitle
\begingroup\renewcommand\thefootnote{\IEEEauthorrefmark{1}}
\footnotetext{Corresponding author.}
\endgroup

\begin{abstract}

Modern instruction-tuned large language models (LLMs) have made remarkable progress in code generation. However, these LLMs fine-tuned with standard supervised fine-tuning (SFT) sometimes generate plausible-looking but functionally incorrect code variants. This issue likely stems from the limitation of standard SFT, which treats all tokens equally during optimization and fails to emphasize the error-sensitive segments—specific code differences between correct implementations and similar incorrect variants. To address this problem, we propose \underline{F}ault-\underline{G}uided F\underline{i}ne-\underline{T}uning (\appname), a novel fine-tuning technique that enhances LLMs' code generation by (1) extracting multi-granularity (line/token-level) differences between correct and incorrect yet similar implementations to identify error-sensitive segments, and (2) dynamically prioritizing those segments during training via dynamic loss weighting. Through extensive experiments on seven LLMs across three widely-used benchmarks, our method achieves an average relative improvement of 6.9\% on pass@1, with some enhanced 6.7B LLMs outperforming closed-source models, e.g., GPT-3.5-Turbo. Furthermore, our fine-tuning technique demonstrates strong generalization with performance improvements ranging from 3.8\% to 19.1\% across diverse instruction-tuned LLMs, and our ablation studies confirm the contributions of different granularities of differences and hyperparameters.
\end{abstract}

\begin{IEEEkeywords}
Large Language Model, Code Generation, Software Engineering
\end{IEEEkeywords}

\maketitle

\input{latex/intro}

\input{latex/Method}

\input{latex/Exp_Setup}

\input{latex/Exp}

\input{latex/Dis}

\input{latex/Threat}

\input{latex/Related}

\input{latex/Conclusion}

\balance
\bibliographystyle{IEEEtran}

\bibliography{IEEEexample}

\end{document}

%% file: latex/intro.tex
\section{Introduction}
Recently, fine-tuning LLMs using synthetic datasets generated by teacher models has emerged as a popular paradigm for improving code generation capabilities~\cite{luowizardcoder, zheng2024opencodeinterpreter, yu2024wavecoder}. This paradigm uses teacher models to generate high-quality instruction-response pairs and construct a dataset. These datasets are then used to fine-tune student models with standard SFT method, which uses instructions to guide LLMs to generate outputs matching reference responses by minimizing cross-entropy loss uniformly across all tokens.

\begin{figure}
    \centering
    \includegraphics[width=0.9\linewidth]{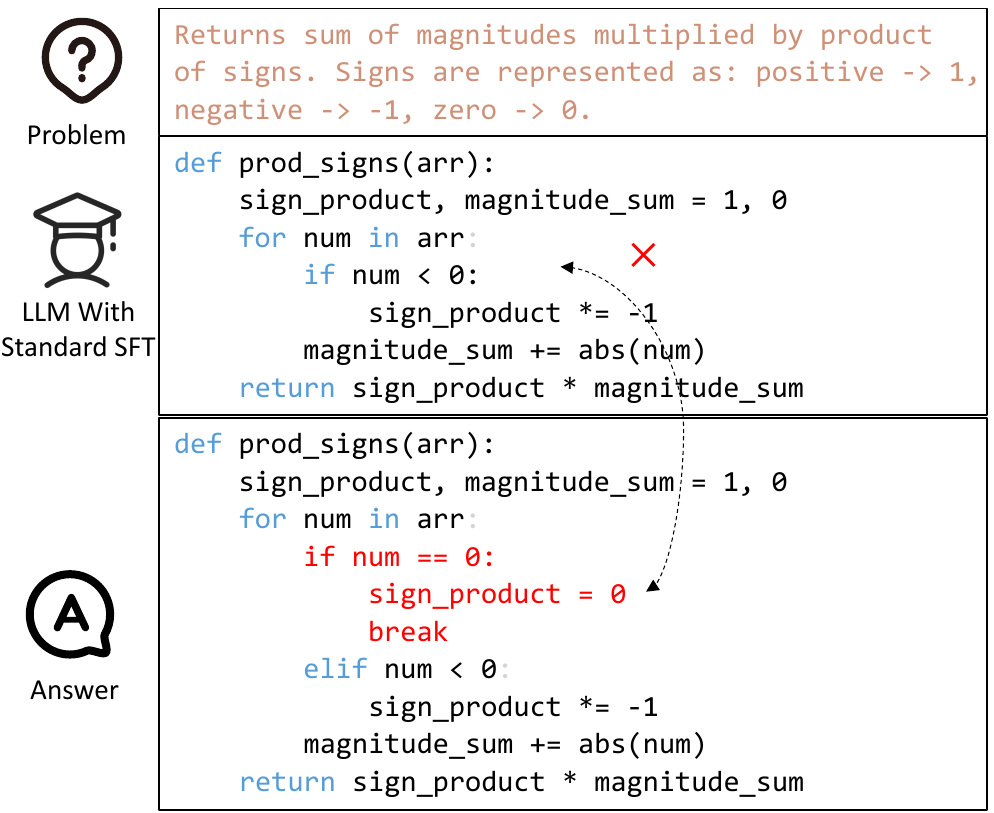}
    \caption{Llama-3.1-70B-Instruct sometimes makes mistakes in \ess in the outputs.}
    \label{fig:motivating}
    \vspace{-0.3cm}
\end{figure}

Although these LLMs fine-tuned with standard SFT achieve impressive performance on code generation benchmarks~\cite{chen2024b4,gunasekar2023textbooks,chen2024jumpcoder,wei2024magicoder}, such as HumanEval~\cite{chen2021evaluating}, \textit{they sometimes generate plausible-looking but incorrect code variants}~\cite{huang2025survey,liu2024exploring}.  
For example, Llama-3.1-70B-Instruct~\cite{dubey2024llama}, a model fine-tuned from Llama-3.1-70B using standard SFT, is capable of solving 82.3\% of the problems in HumanEval. 
However, our analysis shows that 34.5\% of its failed cases are largely correct and require modifications in only three or fewer locations to be fixed. We conservatively classify these specific instances as failures attributable to deviations in error-prone segments, highlighting the prevalence of such errors in code generation. 
As shown in Figure~\ref{fig:motivating}, in a task to calculate the product of signs and the sum of magnitudes, although the LLM successfully implements logic similar to the correct version—by correctly handling negative numbers and summing magnitudes-it fails to account for zero. We refer to such crucial differences between correct implementations and similar incorrect variants as \ess. These \ess act as critical decision points in code generation, where even slight deviations can determine the correctness of the output.

It is crucial to address these errors for code generation. However, standard SFT~\cite{wang2023self} often overlooks this issue as it does not specifically focus on these segments essential for correctness. Drawing inspiration from curriculum learning~\cite{bengio2009curriculum}, which involves training models by progressively moving from easier to more challenging samples to build learning capabilities, we propose \underline{F}ault-\underline{G}uided F\underline{i}ne-\underline{T}uning (\appname). \appname is a supplementary fine-tuning technique designed to guide instruction-tuned models to further focus their learning on these \ess, treating these segments as more challenging samples in a targeted curriculum.

Implementing this approach involves two main challenges. 
The first challenge is to identify these \ess. Systematically collecting large-scale datasets of fine-grained coding errors from real-world scenarios, along with their corresponding correct versions, is difficult.  Furthermore, existing instruction-tuning dataset construction methods primarily focus on generating instruction-response pairs without specifically considering these critical segments~\cite{zhang2023instruction}. Considering traditional code mutation techniques~\cite{koza1994genetic,sun2016finding} are often constrained by a limited set of predefined transformation rules and may not generate diverse and semantically plausible incorrect variants, we develop a two-phase segment identification component. First, we leverage a teacher model, which can capture a wide range of diverse and semantically plausible error patterns from its extensive code training data, with a carefully designed prompt to generate multiple functionally incorrect yet similar variants of the correct implementations from the existing instruction tuning dataset. From these generated variants, we select the one with the most similarity to the correct implementation and highlight the differences as \ess in this paper. 
We then annotate the tokens in the segments through a multi-granularity method at both line and token levels. Notably, this annotation is applied to both the correct implementations and the selected incorrect variant.
The second challenge lies in guiding LLMs to focus on these labeled segments during fine-tuning, as standard SFT treats all tokens with equal importance regardless of their criticality.
To address this challenge, we adjust the loss of SFT to prioritize the annotated error-sensitive tokens within correct implementations. 
Specifically, \appname processes both correct and incorrect implementations to discriminate the \ess, and computes loss based on correct code implementations. Unlike standard SFT, which uniformly weights all tokens during loss computation, we dynamically assign relatively higher weights to those tokens in the correct implementation that correspond to the \ess. This methodology enhances LLMs' capability to discriminate \ess when solving programming tasks, thereby increasing the likelihood of generating correct implementation details while suppressing error-prone alternatives.

To implement our method, we construct a refined dataset derived from the original instruction-tuning data. Each data point consists of an instruction, its correct implementation from the original dataset, and an LLM-generated similar incorrect variant. We then develop a multi-granularity error-sensitive segment extraction method and combine it with the refined loss function to enhance LLM's code generation capabilities.

We validate the effectiveness of the \appname through extensive experiments. Notably, through \appname, the selected LLMs achieve an average relative improvement of 6.9\% on pass@1 across three representative code generation benchmarks (HumanEval(+), MBPP(+), and BigCodeBench)~\cite{chen2021evaluating,austin2021program,liu2024your,zhuo2024bigcodebench}. Among these LLMs, SemCoder-S~\cite{dingsemcoder} with 6.7B parameters outperforms closed-source models like GPT-3.5-Turbo~\cite{OpenAI2022} on HumanEval(+) and MBPP(+) benchmarks, and MagiCoder$\mathcal{S}$-DS with 6.7B parameters outperforms GPT-3.5-Turbo on HumanEval(+). Our method also demonstrates strong generalization capabilities, showing performance improvements ranging from 3.8\% to 19.1\% across multiple instruction-tuned LLMs, including those trained on closed-source instruction datasets. Moreover, our ablation experiments on \appname confirm the contributions of different granularities of differences and hyperparameters.

\noindent We summarize our contributions as follows.
\begin{itemize}
   \item To the best of our knowledge, we are the first to investigate how to enhance LLMs' understanding of \ess by refining the SFT process to improve LLMs' code generation capabilities.
   \item We propose a novel framework, Fault-Guided Fine-Tuning (\appname), to effectively guide LLMs to focus on error-prone parts in code. This is achieved by (1) extracting multi-granularity code differences (token-/line-level) to identify \ess, and (2) refining SFT to dynamically assign higher weights to these parts during the training process.
   \item Through extensive experiments across seven LLMs and three widely-used code generation benchmarks, we demonstrate the effectiveness and generalizability of our approach in effectively boosting LLMs' code generation performance compared to baseline methods.
\end{itemize}

%% file: latex/Method.tex
\section{Approach}
\label{sec:3}
\begin{figure*}
    \centering
    \includegraphics[width=0.9\linewidth]{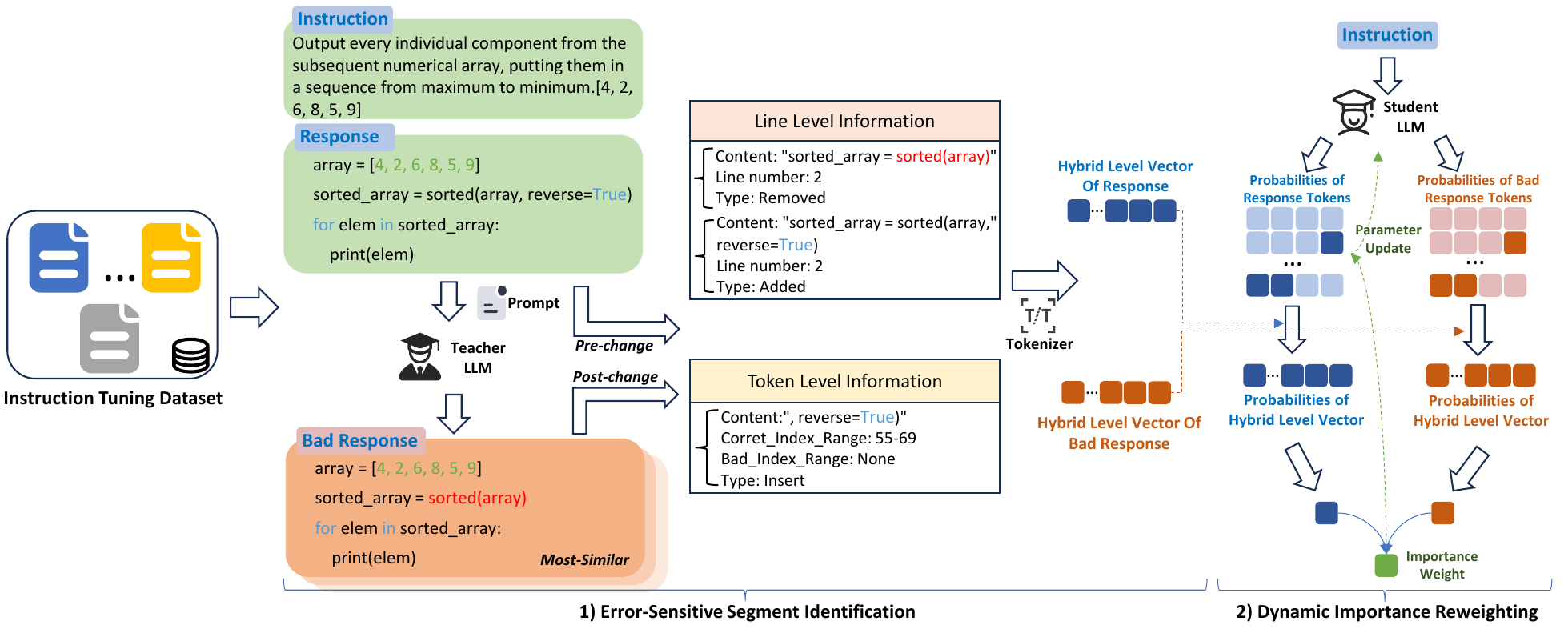}
    \caption{The overview of \appname, taking one sample for explanation.}
    \label{fig:overview}
    \vspace{-0.5cm}
\end{figure*}
Figure~\ref{fig:overview} illustrates the overview of \appname. This approach takes as input an instruction-tuned LLM and its corresponding instruction-tuning dataset, and outputs an enhanced LLM with improved code generation capabilities that can better discriminate \ess. It first augments the dataset by generating similar yet incorrect implementations for each correct response. Then, it identifies \ess between the paired implementations and calculates weights for tokens in the correct implementations that differ from the incorrect variants. During fine-tuning, it only computes loss on the correct implementations, with higher weights assigned to tokens in \ess, producing an LLM that can better discriminate these \ess.  The methodology consists of two key components:

\begin{enumerate}
\item \textit{Error-Sensitive Segments Identification}:  This component creates a refined dataset of paired correct and similar yet wrong code samples from the original dataset, and processes code differences at multiple granularities to identify \ess.
\item \textit{Dynamic Importance Reweighting}: This component strategically reweights token weights in the loss function to prioritize discriminative elements in correct implementations, building upon the identified \ess. This dynamic weighting method enhances the LLM's attention to the key implementation details in correct code, effectively teaching it to distinguish between valid solutions and their similar yet incorrect counterparts, ultimately improving code generation capabilities.
\end{enumerate}

The two components work together to fine-tune LLMs to distinguish between correct implementations and similar yet incorrect alternatives, thereby improving performance. 

\subsection{Error-Sensitive Segments Identification}
\begin{figure}[t]
    \centering
    \includegraphics[width=0.85\linewidth]{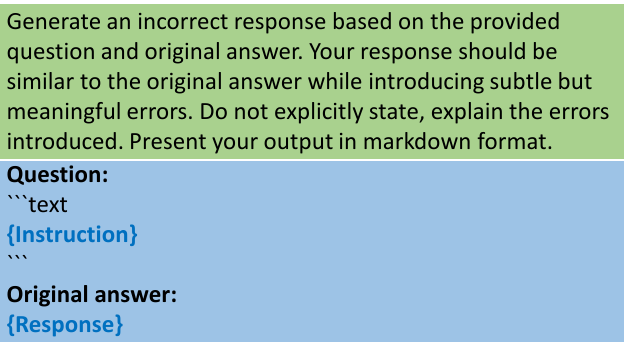}
    \caption{The prompt for generating similar yet incorrect response.}
    \label{fig:prompt}
    \vspace{-0.2cm}
\end{figure}

The input to this component is the instruction-tuning dataset $\mathcal{D} = {(c^{\text{correct}}_i, p^{\text{target}}_i)}_{i=1}^{N}$, where $p^{\text{target}}_i$ represents the target problem description and $c^{\text{correct}}_i$ denotes the correct implementation. The output is an enhanced dataset $\mathcal{D} = {(c^{\text{correct}}_i, c^{\text{incorrect}}_i, p^{\text{target}}_i)}_{i=1}^{N}$ with error-sensitive segment information, where $c^{\text{incorrect}}_i$ represents the corresponding similar but incorrect implementation. To generate incorrect code variants, we utilize a teacher LLM with a carefully designed prompt. We choose an LLM-based approach over code mutation techniques because teacher models, by leveraging error commonalities learned from their extensive code corpora, offer better flexibility in producing a diverse range of incorrect variants. The prompt template is shown in Figure~\ref{fig:prompt}, which consists of two parts. The first part defines the task for producing incorrect responses corresponding to the target problem and answer, specifying that outputs should be similar to correct solutions, with responses constrained to markdown formatting for consistent post-processing. The second part provides contextual references to the target problem description and solution.

We generate multiple incorrect variants for each correct implementation and select the one that is most similar to the correct solution. To identify the most similar one, we employ an embedding-based approach because embeddings can capture semantic similarities that purely lexical comparisons might overlook. Specifically, we generate embeddings with UnixCoder~\cite{guo2022unixcoder} for its deep understanding of code structures and semantics derived from its diverse, large-scale code corpora. We then extract the differences to identify \ess and process them at different granularity levels to capture both line-level and token-level information. Specifically, we designate $c^{\text{incorrect}}$ as the \emph{pre-change} version and $c^{\text{correct}}$ as the \emph{post-change} version.

\textit{Line-Level Differences}. We align $c^{\text{correct}}_i$ and $c^{\text{incorrect}}_i$ line-by-line using Python's \texttt{difflib} library. For each line, it assigns a flag indicating whether it should be deleted (\texttt{-}), added (\texttt{+}), or remain unchanged. We extract the lines marked for deletion from $c^{\text{incorrect}}$ and those marked for addition from $c^{\text{correct}}$.

Let $L_c$ and $L_a$ denote the number of code lines in the correct code $c^{\text{correct}}$ and incorrect code $c^{\text{incorrect}}$, respectively. Based on these extracted lines, we construct the line-level boolean mask vectors $V_{\text{line}}^c$ and $V_{\text{line}}^a$ for $c^{correct}$ and $c^{incorrect}$ as follows:
\begin{small} 
\begin{align*}
V_{\text{line}}^c &= [v^c_1, v^c_2, \dots, v^c_{L_c}],  \text{where } v^c_i = I(\text{line}_i^c \text{ is added}) \\
V_{\text{line}}^a &= [v^a_1, v^a_2, \dots, v^a_{L_a}],  \text{where } v^a_j = I(\text{line}_j^a \text{ is deleted})
\end{align*}
\end{small}
where $I(\cdot)$ is the indicator function that outputs 1 if the condition is true and 0 otherwise.

\textit{Token-Level Differences}. We utilize the Levenshtein distance algorithm~\cite{yujian2007normalized} to identify character-level change information between $c^{\text{incorrect}}$ and $c^{\text{correct}}$. The Levenshtein distance algorithm, also known as the edit distance algorithm, quantifies the minimum number of single-character operations (insertions, deletions, or substitutions) required to transform one string into another. We identify the characters that need to be edited to transform the original string ($c^{\text{incorrect}}_i$) into the modified version ($c^{\text{correct}}$), and record their positions accordingly. For instance, if a character operation is an insertion, we record its position in $c^{\text{correct}}$, as shown in Figure~\ref{fig:overview}. Given that the LLM's embedding layer is tightly coupled with the LLM's tokenizer vocabulary, we map these character-level differences to tokens using the LLM's tokenizer. When character modifications span multiple tokens, all affected tokens are marked.

Let $T_c$ and $T_a$ denote the number of tokens in $c^{\text{correct}}$ and $c^{\text{incorrect}}$. We construct token-level boolean mask vectors $V_{\text{token}}^c$ and $V_{\text{token}}^a$ for $c^{correct}$ and $c^{incorrect}$ as follows:
\begin{small}
\begin{align*}
V_{\text{token}}^c &= [w^c_1, \dots, w^c_{T_c}], \; w^c_k = I(\text{token}_k^c \text{ is added}) \\
V_{\text{token}}^a &= [w^a_1, \dots, w^a_{T_a}], \; w^a_\ell = I(\text{token}_\ell^a \text{ is deleted})
\end{align*}
\end{small}

\textit{Hybrid Level Vectors}.
To create comprehensive representations of \ess, we combine line-level and token-level masks. However, these two types of masks operate at different granularities and cannot be directly combined. To get line-level masks with token-level granularity, we first initialize a new token-level mask vector with all zeros, corresponding to the total number of tokens in the code. Then, for each line in the original code, all tokens belonging to that line are assigned the line's mask value (1 if the line is marked, 0 if it is unmarked) in this new token-level mask. This process yields the token-level representations $V_{\text{line-to-token}}^c$ and $V_{\text{line-to-token}}^a$ when applied to $V_{\text{line}}^c$ and $V_{\text{line}}^a$ respectively.
We then use an element-wise addition operation to combine $V_{\text{line-to-token}}$ and $V_{\text{token}}$, as follows:
\begin{align*}
V_{\text{hybrid}}^c &= V_{\text{line-to-token}}^c +  V_{\text{token}}^c \\
V_{\text{hybrid}}^a &= V_{\text{line-to-token}}^a + V_{\text{token}}^a
\end{align*}

These hybrid vectors precisely identify \ess at multiple granularities, highlighting critical differences between correct and incorrect implementations. Noted that changed tokens must appear in changed lines, our hybrid representation naturally creates a priority system: 1) tokens that are both in changed lines and are themselves changed will have a value of 2 in the hybrid vector; 2) tokens that are only in changed lines but not directly changed will have a value of 1. This provides a more comprehensive view than either granularity alone, with higher values indicating more critical tokens.

\subsection{Dynamic Importance Reweighting}
With the identified \ess, we now refine the SFT process to prioritize these critical differences. Based on the constructed dataset $\mathcal{D} = \{(c^{\text{correct}}_i,c^{\text{incorrect}}_i,p^{target}_i)\}_{i=1}^{N}$, the standard SFT loss is computed as:
\begin{equation}
\label{equ:sft}
\begin{aligned}
\mathcal{L}_{SFT} = -\frac{1}{n}\sum_{i=1}^{N}{\sum_{j=1}^{T_c}logP(c^{\text{correct}}_{i,j}|p^{target}_i,c^{\text{correct}}_{i,1:j-1})}
\end{aligned}
\end{equation}

\noindent where $N$  denotes the number of samples in a batch. Notably, the standard SFT loss function treats all tokens equally.

In contrast, \appname introduces dynamic token-level weights $W={w_1,w_2,...,w_j}$ emphasize \ess:
\begin{equation}
\label{equ:fault}
\begin{aligned}
\mathcal{L}_{Fault} = -\frac{1}{n}\sum_{i=1}^{N}{\sum_{j=1}^{T_c} w_j \cdot  logP(c^{\text{correct}}_{i,j}|p^{target}_i,c^{\text{correct}}_{i,1:j-1})}
\end{aligned}
\end{equation}
The weight $W$ is computed as follows: Given input $x = p^{\text{target}}$, outputs $y^c = c^{correct}$ and $y^a=c^{incorrect}$,  we first obtain the LLM's prediction probabilities for both correct and incorrect implementations given the same instruction:

\begin{align}
P^c &= f_{\theta}( y^c_{k} \mid y^c_{1:k-1}, x ) \\
P^a &= f_{\theta}( y^a_{l} \mid y^a_{1:l-1}, x )
\end{align}
where $f_{\theta}$ represents the conditional probability function of the LLM that computes the probability of the next token given the input $x$ and previous tokens.
We then apply the hybrid-level vectors to isolate probabilities for \ess:

\begin{align}
H^c &= P^c \odot V_{\text{hybrid}}^c \\
H^a &= P^a \odot V_{\text{hybrid}}^a
\end{align}
Where $\odot$ denotes element-wise multiplication. Inspired by the Bradley–Terry model~\cite{hunter2004mm}, a pairwise comparison framework widely used in ranking systems~\cite{baker2021modifying,yan2016ranking,menke2008bradley}, we compute dynamic token weights $W$ for differentiating tokens in \ess:

\begin{small}
\begin{align*}
    W = \alpha \ - \vert \frac{\overline{H^c}-\overline{H^a}}{\overline{H^c}+\overline{H^a}} \vert 
\end{align*}
\end{small}
where $\alpha$ is a hyperparameter controlling the weight range, $\overline{H^c}$ denotes the mean probability of tokens in \ess in $c^{\text{correct}}_i$, and $\overline{H^a}$ represents the corresponding value for $c^{\text{incorrect}}$. This formulation ensures that: (1) When the mean probabilities $\overline{H^c}$ and $\overline{H^a}$ are close (indicating the LLM struggles to distinguish the tokens between $c^{\text{correct}}$ and $c^{\text{incorrect}}$), the weights for differentiating tokens in $c^{\text{correct}}$ approach $\alpha$, thereby maximizing emphasis on \ess. (2) Conversely, when $\overline{H^c}$ and $\overline{H^a}$ diverge significantly (demonstrating the LLM can discriminate the differentiating tokens in $c^{\text{correct}}$ and $c^{\text{incorrect}}$), the weights diminish toward $\alpha-1$, reducing emphasis. For tokens shared between $c^{\text{correct}}$ and $c^{\text{incorrect}}$, we assign fixed weights $\alpha-1$, ensuring the LLM maintains baseline attention to shared elements while prioritizing discriminative features. To be noted that this dynamic weight $W$ is differentiable, as $\overline{H^c}$ and $\overline{H^a}$ are derived from the model's output probabilities for $c^{\text{correct}}$ and $c^{\text{incorrect}}$ within the \ess. Consequently, through gradient updates, the optimization process reinforces $c^{\text{correct}}$, also implicitly steering the model away from generating $c^{\text{incorrect}}$.

This dynamic weighting mechanism guides the LLM to focus on challenging discriminative aspects of correct implementations, which can improve its code generation capability.

%% file: latex/Exp_Setup.tex
\section{Experiments Setup}
\label{sec:4}

\subsection{Benchmarks and Metrics}
We conduct experiments on three widely used code generation benchmarks to demonstrate the superiority and generality of \appname. We use HumanEval(+),  MBPP(+)~\cite{liu2024your,chen2021evaluating,austin2021program} and Bigcodebench\cite{zhuo2024bigcodebench}. Specifically, our study uses both the full set and hard set with complete configuration of Bigcodebench, namely, BigCodeBench-Full and BigCodeBench-Hard. 

To evaluate performance, we use the Pass@K metric, which is widely used in prior studies~\cite{chen2021evaluating,mu2023clarifygpt,huang2023competition}. Following prior studies~\cite{fakhoury2024llm,dong2024self,jiang2024self}, our experimental design adopts K=1, focusing exclusively on first-attempt success rates. This metric also aligns with real-world scenarios where developers aim to produce accurate code on the first attempt~\cite{dong2024self}.

\subsection{Implementation Detail} 

\subsubsection{Data generation} We use Qwen2.5-Coder-32B-Instruct~\cite{hui2024qwen2} as the teacher model with temperature=0.8 to generate three incorrect code implementations for each sample. 
Our rationale for selecting this model is threefold. First, generating an incorrect solution by referencing a correct one is comparatively less demanding on a model's capabilities than generating a correct solution from scratch. Secondly, our task benefits from a model with strong coding abilities, and Qwen2.5-Coder-32B-Instruct possesses high coding proficiency; for instance, its performance on HumanEval(+) and MBPP(+) even surpasses that of GPT-4o. Lastly, using open-source models also facilitates reproducibility.

To investigate whether the teacher model can generate implementations that are incorrect and similar to the correct solutions, we employ both manual inspection and automated quantitative validation. First, two authors independently examine a sample of 50 generated outputs, which are randomly sampled from the Evol-instruct dataset, assessing them based on two criteria: functional correctness and similarity to the original correct code. A discussion is held to resolve the disagreements. We do not invite others because all the disagreements are resolved during discussion. The Cohen’s Kappa coefficient~\cite{mchugh2012interrater} is 0.78. Our analysis shows that the LLM could produce similar yet incorrect samples as expected: 94\% of the generated samples are similar to the correct ones yet incorrect, while the remaining 6\% introduce changes that do not affect correctness. To quantitatively analyze the correctness and similarity of the generated responses, we select benchmarks with test cases for further validation. Specifically, we choose HumanEval and MBPP to generate incorrect yet similar responses and use Unixcoder\cite{guo2022unixcoder} to calculate cosine similarity. We find that 94.1\% of generated responses are incorrect using the test cases in benchmarks, and the average similarity score is 0.95, aligning with our needs. Due to space limitations, the manually checked samples and generated responses are in our replication package.

\subsubsection{Settings}
All experiments are conducted on a machine with eight Tesla A800 GPUs, each with 80 GB of memory. $\alpha$ is set to 2, which means the weight range of $W$ is (1,2). 
Given that \appname is designed for already instruction-tuned LLMs, where the objective is not to train them extensively from a base state but rather to refine their existing capabilities to better handle \ess present, we train all models for one epoch with a relatively low learning rate of 5e-6, aiming to gently adapt the model and prevent catastrophic forgetting of previously learned knowledge, while efficiently instilling the new focus on these critical segments~\cite{choi2025teaching,ortiz2025gender}.
The max sequence length is 1024. For inference evaluation, we use greedy decoding to ensure deterministic outputs, which also aligns with prior studies~\cite{chen2021evaluating,mu2023clarifygpt}.

%% file: latex/Exp.tex
\section{Results}
\label{sec:5}

In this section, we report and analyze the experimental results to answer the following research questions (RQs):
\begin{itemize}[leftmargin=*]
\item RQ1: How effective is our approach in improving code generation across different benchmarks?
\item RQ2: How do different components of the \appname method contribute to LLMs' performance?
\item RQ3: Does \appname demonstrate generalizability across different LLMs and their corresponding instruction-tuning datasets?
\item RQ4: Does \appname work for instruction-tuned LLMs whose instruction-tuning dataset is closed-source?
\end{itemize}

\subsection{RQ1: Overall Effectiveness}
In this RQ, we evaluate the effectiveness of our approach by applying it to several Instruction-tuned LLMs using their corresponding instruction-tuning datasets and assess their performance against four baselines:\\
\textit{Base Models:} We use the original instruction-tuned LLMs without any additional \appname as our base models. This comparison demonstrates the absolute improvement achieved through \appname. Specifically, we select three representative instruction-tuned LLMs: MagiCoder$\mathcal{S}$-CL~\cite{wei2024magicoder}, MagiCoder$\mathcal{S}$-DS~\cite{wei2024magicoder} and SemCoder-S~\cite{dingsemcoder} as our base models.\\
\textit{Closed-Source Model:} We include GPT-3.5-Turbo~\cite{OpenAI2022} as the closed-source baseline to illustrate the performance gap between our fault-guided fine-tuned LLMs and the advanced closed-source LLM.\\
\textit{Standard-SFT Models:} We apply standard SFT on the same base models to create this baseline. This comparison serves two purposes: (1) to examine whether further fine-tuning on coarse-grained instruction-response mappings on their existing dataset can improve performance over the original models, and (2) to highlight the superior performance of our approach in learning fine-grained \ess.\\
\textit{Other Open Source Instruction-Tuned Code LLMs:}  We include other instruction-tuned versions of the same base models, which are trained on additional data generated by stronger LLMs (e.g., GPT-4) in addition to Evol-Instruct. Specifically, we include WaveCoder-Ultra~\cite{yu2024wavecoder}, OpenCodeInterpreter\cite{zheng2024opencodeinterpreter}, 
and AlchemistCoder~\cite{song2024alchemistcoder}.
This comparison aims to illustrate how the performance of the selected LLMs with \appname compares to that of current popular instruction-tuned models based on the same foundational model.

We choose Evol-instruct~\cite{wei2024magicoder}, a decontaminated version of evol-codealpaca-v1~\cite{evolalpaca}, which contains numerous high-quality instruction-following data, as our training dataset. This dataset is decontaminated by removing data that contain docstrings or solutions from multiple benchmarks~\cite{chen2021evaluating,austin2021program,cassano2023multipl,lai2023ds}. We conduct an additional decontamination of this dataset for docstrings or solutions from BigCodeBench following~\cite{wei2024magicoder}, and find no overlap. For MagiCoder$\mathcal{S}$-CL and MagiCoder$\mathcal{S}$-DS, this dataset is their original instruction-tuning dataset. For SemCoder-S, this dataset is a subset of its original instruction-tuned dataset, which is not fully open-sourced.

\input{Table/main-all}
We report results consistently from the EvalPlus leaderboard\footnote{https://evalplus.github.io/leaderboard.html} and BigCodeBench leaderboard\footnote{https://bigcode-bench.github.io}.
Table~\ref{tab:main-all} presents the performance of LLMs with \appname and the baselines across HumanEval(+), MBPP(+), and BigCodeBench. Overall, LLMs with \appname demonstrate substantial improvements in code generation. We observe that LLMs with \appname show average relative performance improvements of 4.8\% over the base model and 4.9\% over LLMs with Standard-SFT on HumanEval(+), MBPP(+), and BigCodeBench. Notably, with \appname, SemCoder-S with only 7B parameters outperforms  GPT-3.5-Turbo on HumanEval(+) and MBPP(+), and MagiCoder$\mathcal{S}$-DS outperforms GPT-3.5-Turbo on HumanEval(+). Both LLMs achieve comparable performance to GPT-3.5-Turbo on BigCodeBench, further validating the exceptional effectiveness of \appname in enhancing code generation capabilities. While the improvement on BigCodeBench-Full is modest, our approach shows more gains on BigCodeBench-Hard (e.g., 20.8\% relative improvement for SemCoder-S). This is likely because more challenging problems contain more \ess, and \appname is designed to guide LLMs to handle these \ess, thus showing greater effectiveness on difficult programming tasks.

When comparing the effects of \appname versus Standard-SFT on base models, we observe that Standard-SFT provides limited improvements and sometimes even weakens the base models. For example, SemCoder-S with Standard-SFT achieves only 0.8\% relative performance improvement on HumanEval(+) and suffers 1.9\% relative performance decline on MBPP+. This suggests that simply reinforcing the coarse-grained instruction-response mappings on their existing dataset provides minimal benefits, as these models have already captured these general mappings well during their initial instruction tuning. When compared with other instruction-tuned models, we observe that by enabling models to further learn the \ess within the original dataset through \appname, their performance can achieve comparable or even superior results.

\begin{figure}
    \centering
    \includegraphics[width=0.85\linewidth]{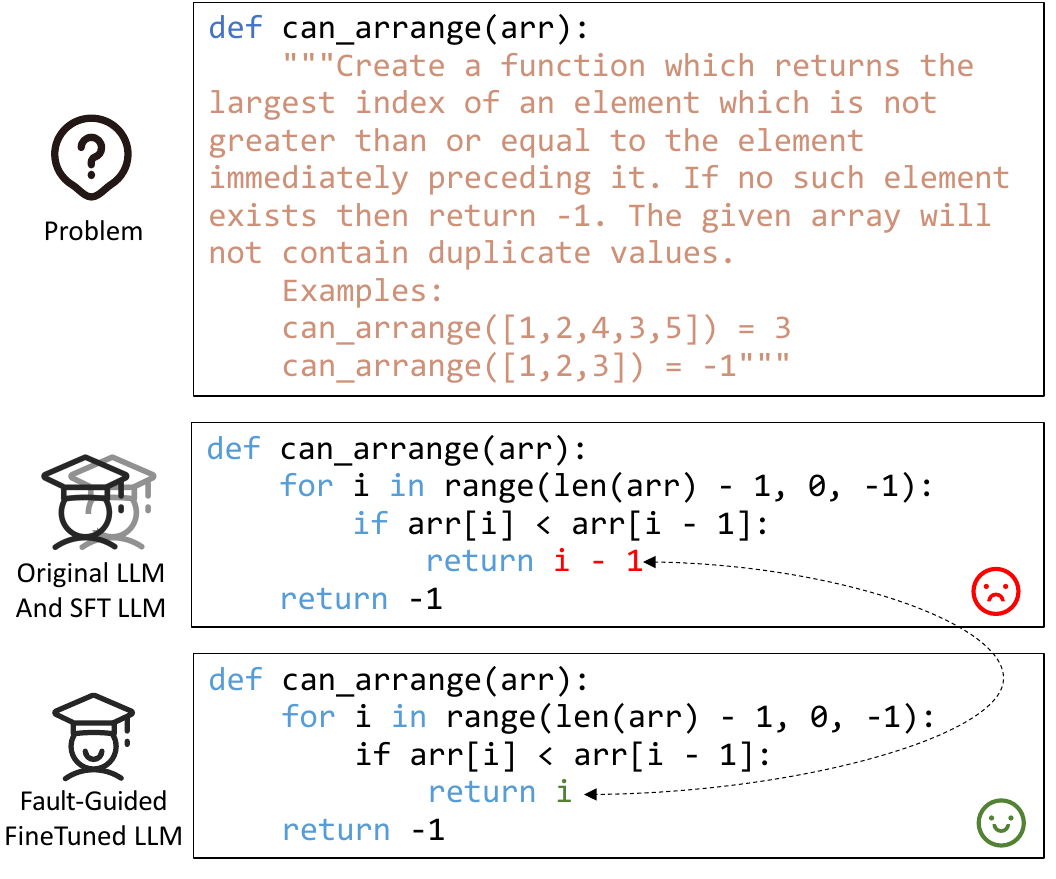}
    \caption{A case demonstrating how LLMs after \appname can better focus on \ess to generate the correct solution.}
    \label{fig:dis-case}
    \vspace{-0.5cm}
\end{figure}

To further figure out the reasons for \appname improving LLMs' ability to generate functionally correct code, we manually inspect the results. Based on our analysis, \appname demonstrates two main advantages over both the original model and Standard-SFT:

First, \appname can learn diverse \ess to better recognize and focus on implementation details that are prone to errors, while the original model and Standard-SFT only learn the overall mapping from problem to solution. This method improves the LLM's attention to key implementation choices. 
We manually examine the tasks from HumanEval(+) that are correctly solved by applying \appname but are incorrect when applying Standard-SFT. We find that for SemCoder-S, \appname correct 11 tasks where Standard-SFT failed. In 81.8\% of these instances, Standard-SFT’s failure stemmed from incorrectly handling \ess; these errors typically required modifications of three lines of code or fewer to be rectified. \appname, in contrast, provides a correct implementation, with similar rates observed in MagiCoder$\mathcal{S}$-DS (71.4\%) and MagiCoder$\mathcal{S}$-CL (66.7\%). 
For example, Figure~\ref{fig:dis-case} presents a comparison of the results of three versions of SemCoder-S on the HumanEval/135 task. In this example, an error-sensitive segment involves deciding whether to return the index of the target element itself \textit{i} or the index of its previous element \textit{i-1}. This distinction directly impacts the functional correctness of the implementation. Both the original model and the model with Standard-SFT incorrectly return the index of the previous element, while the model with \appname correctly returns the index of the target element. These results further confirm our method's effectiveness in guiding LLMs to recognize \ess.

\begin{figure}
    \centering
    \includegraphics[width=0.85\linewidth]{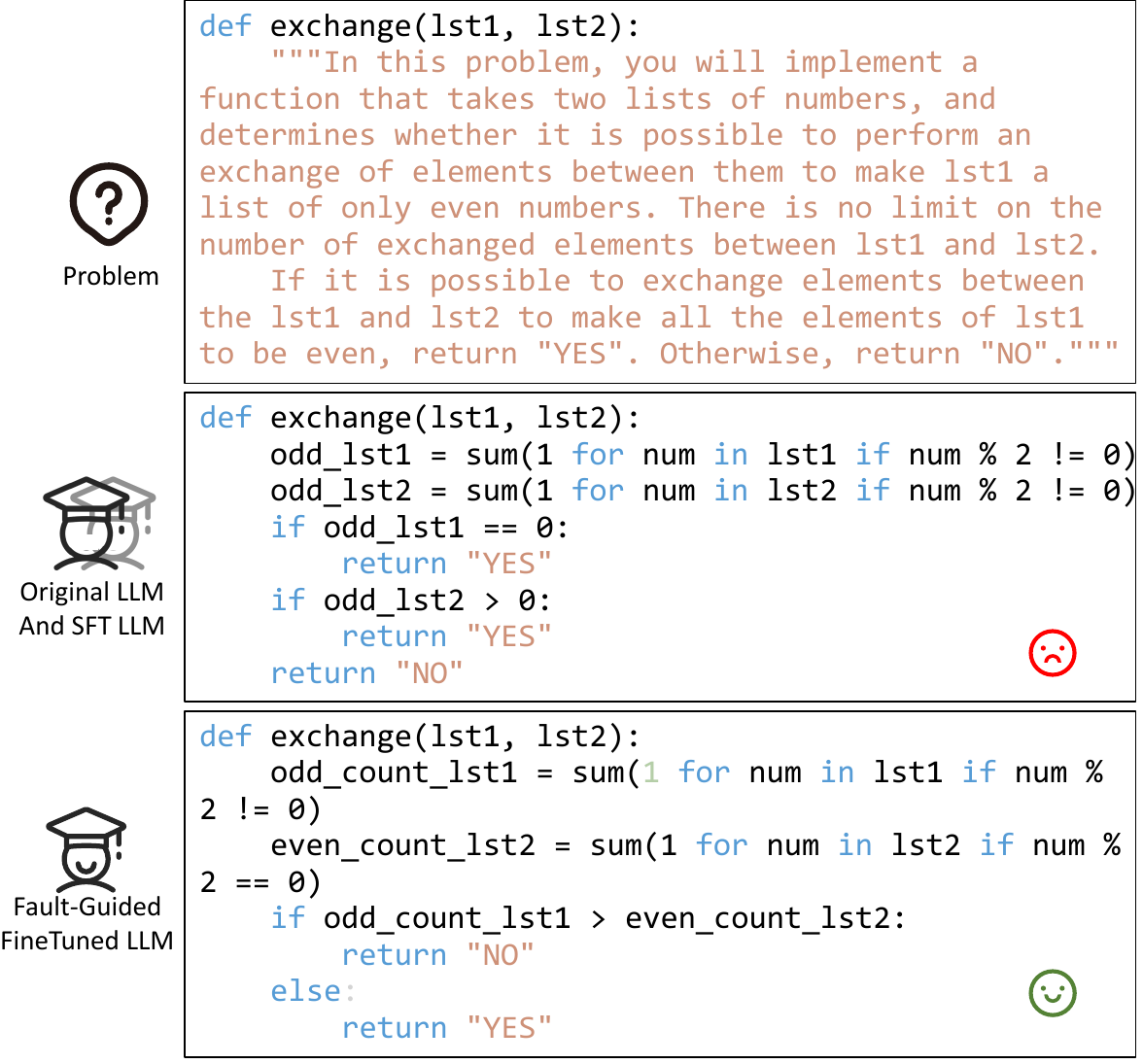}
    \caption{A case demonstrating how \appname can improve overall code generation performance.}
    \label{fig:dis-case-2}
    \vspace{-0.1cm}
\end{figure}

Second, by developing a deeper understanding of critical code segments, \appname also enhances overall code generation capabilities. By strategically emphasizing \ess while maintaining appropriate weight for contextual elements, the LLM learns to identify and handle the crucial parts of implementations that determine correctness. 
Figure~\ref{fig:dis-case-2} demonstrates this using an example from SemCoder-S on HumanEval/110, which requires determining whether swapping elements between two lists can make all elements in \textit{lst1} even. 
This case highlights improvements that go beyond addressing specific \ess.
Both the original model and Standard-SFT fail to implement the correct verification logic to determine whether \textit{lst2} contains enough even numbers to replace odd numbers in \textit{lst1}. 
In contrast, the Fault-Guided Fine-Tuned model correctly implements this logic, demonstrating enhanced general coding abilities rather than just handling error-sensitive parts.

\begin{center}
    \begin{mybox} \textbf{RQ1 Summary: } %
    \appname delivers consistent and substantial performance improvements across all three benchmarks, with enhanced LLMs even outperforming GPT-3.5-Turbo on certain benchmarks. The results confirm that explicitly learning fine-grained error-sensitive segment mappings is more effective than simply retraining on coarse-grained instruction-response pairs.
    \end{mybox} 
\end{center}
\subsection{RQ2: Component Analysis}
\input{Table/abl-evalplus}
To understand how different components contribute to the effectiveness of \appname, we conduct ablation studies focusing on the impacts of multi-granularity and the loss function. 

\textit{Impact of Difference Granularity.} We explore the impact of code difference granularity, which involves synthesizing line-level and token-level code differences to identify \ess. Specifically, we conduct ablation experiments using MagiCoder$\mathcal{S}$-DS and SemCoder-S as base models and perform evaluation on three selected benchmarks. Table~\ref{tab:abl-evalplus} shows the impact of different granularities of differences on \appname.  We can observe that the combination of line-level granularity and token-level granularity yields maximum performance gains. For example, when applied to SemCoder-S, this approach achieves a relative average improvement of 4.9\% across all benchmarks compared to the base model, compared to just 2.9\% for line-level only and 2.4\% for token-level only. 

We further investigate models' ability to handle \ess after combining both granularities, compared to the base models. We select the tasks from HumanEval(+) that are correctly solved after applying \appname but initially incorrect with the base models. We find that among these tasks, 63.6\% of SemCoder-S's improvements result from properly handling \ess, with these specific \ess-related corrections involving modifications of three lines or fewer. Similar rates are observed in MagiCoder$\mathcal{S}$-DS (71.4\%).  This demonstrates that multi-granularity differences enable better \ess, thereby enhancing LLMs' code generation capabilities.

\begin{figure}
    \centering
    \includegraphics[width=0.9\linewidth]{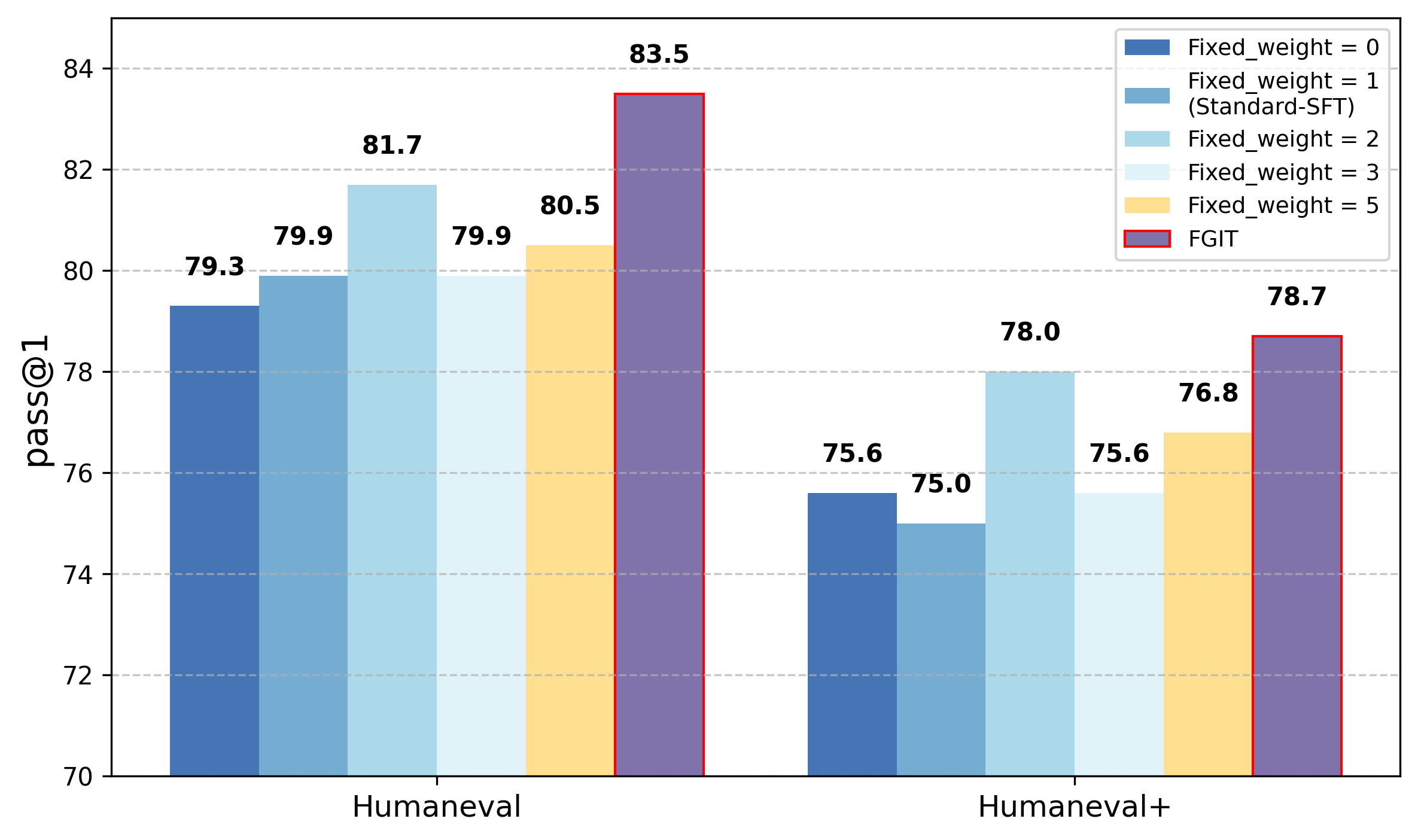}
    \caption{Performance of different weights based on SemCoder-S on HumanEval(+)}
    \label{fig:abl-weight}
    \vspace{-0.5cm}
\end{figure}

\textit{Impact of the Loss Function.} To validate the effectiveness of our dynamic loss weighting design, we compare our dynamic weighting approach against a fixed weighting strategy where all error-sensitive tokens receive the same constant weight during training. We experiment with fixed weights in the range of $[0, 1, 2, 3, 5]$ on HumanEval(+). Due to the evaluation time constraints, we select SemCoder-S as the representative LLM for this ablation study, as it demonstrates the best overall performance with \appname. Notably, when the fixed weight equals 1, this configuration is equivalent to standard SFT. Figure~\ref{fig:abl-weight} shows the performance of SemCoder-S with different fixed weights compared to our dynamic weighting approach. Our experimental results reveal two important findings: (1) A fixed weight of 2 yields better performance than other fixed weight values, suggesting that an appropriate constant weight can help the LLM recognize \ess and enhance code generation capabilities. (2) Our dynamic weighting approach still outperforms the best fixed weighting configuration. This confirms that dynamically adjusting weights based on the LLM's current discrimination ability provides better guidance for the LLM to focus on critical implementation details that differentiate correct solutions from their erroneous variants.

\begin{center}
    \begin{mybox} \textbf{RQ2 Summary: } %
    All components in \appname contribute to the performance. Combining different levels of granularity of code differences (line + token level) is critical to performance. The loss function with dynamic weighting strategies outperforms that with fixed weighting strategies, highlighting the effectiveness of our weighting method. 
    \end{mybox} 
\end{center}

\subsection{RQ3: The Generalization Capabilities}
\input{Table/gen-eval}

In this RQ, we aim to explore the generalizability of \appname across different instruction-tuned LLMs when using their own instruction-tuning datasets. Specifically, we select two representative LLMs and their corresponding instruction datasets. 1)  We select MagiCoder-DS and its corresponding dataset OSS-Instruct. This dataset is generated from open-sourced code by GPT-3.5-Turbo, and contains 75K samples. 2) SemCoder and its corresponding dataset PYX. This dataset consists of 95K samples, including comprehensive reasoning texts with executable code samples. The dataset is constructed with problem descriptions generated by GPT-3.5-Turbo and corresponding responses generated by GPT-4o-mini~\cite{OpenAI2024GPT4o}, creating high-quality instruction-response pairs with detailed reasoning. For each LLM, we process its corresponding dataset through our pipeline and evaluate performance on the same benchmarks used in RQ1. To be noted, we also decontaminate the datasets, adhering to~\cite{wei2024magicoder}, and find no data overlap with our selected evaluation benchmarks. We select Base Models and Standard-SFT Models as our baselines. By using LLMs with different training paradigms and datasets with varying characteristics (open-sourced code versus detailed reasoning with executable samples), we can verify that our approach is not tied to specific LLM series or dataset properties, but rather provides universal benefits.

Table~\ref{tab:gen-eval} shows the performance of LLMs on HumanEval(+), MBPP(+), and BigCodeBench after \appname and Standard-SFT. We can observe that \appname demonstrates robust generalization capabilities to different instruction-tuning LLMs and their corresponding datasets. For MagiCoder-DS and SemCoder, after \appname, the average performances across all benchmarks show relative improvements of 5.3\% and 3.8\% compared to the base models. In contrast, standard SFT yielded modest relative improvements of 1.4\% for MagiCoder-DS and decreased performance by 2.1\% for SemCoder. These results show that \appname's benefits are not tied to specific dataset characteristics or model series. Instead, the approach effectively enhances diverse instruction-tuned LLMs by teaching them to focus on \ess in correct solutions.

\begin{center}
    \begin{mybox} \textbf{RQ3 Summary:}
 \appname exhibits strong generalizability across different instruction-tuned LLMs and their corresponding datasets, consistently outperforming standard SFT. 
    \end{mybox} 
\end{center}

\subsection{RQ4: Effectiveness on LLMs with Closed-Source Instruction Data}
\input{Table/rq4-eval}

A key question for the broader adoption is whether \appname can enhance LLMs whose original instruction-tuning datasets are not publicly available. To investigate this, we applied our method to two widely-used LLMs with closed-source training data in this RQ. Specifically, we choose CodeLlama-7B-Instruct~\cite{roziere2023code} and DeepSeek-\\Coder-6.7B-Instruct~\cite{guo2024deepseek} as base models. These LLMs are instruction-tuned on substantial but proprietary datasets - CodeLlama-7B-Instruct underwent instruction tuning on approximately 5B tokens of instruction data, while DeepSeekCoder-6.7B-Instruct is tuned on around 2B tokens. To test our approach without access to these original datasets, we select Evol-Instruct, used in the main experiment, as the training dataset. We select Base Models, and Base Models with Standard SFT as our baselines and evaluate on HumanEval(+), MBPP(+) and BigCodeBench(+).

Table~\ref{tab:rq4-eval} shows the performance of these two LLMs using Standard-SFT and fault-fine-tuning on HumanEval(+), MBPP(+), and BigCodeBench. We can find that \appname is also applicable to LLMs with closed-source datasets. For CodeLlama-Instruct and DeepseekCoder-Instruct, after \appname, the average relative improvements across all benchmarks are 19.1\% and 5.9\%. By comparison, the standard SFT yield gains of 7.5\% and 0.5\%, respectively. This further demonstrates that fault-fine-tuning is also applicable to LLMs with closed-source datasets, showing strong applicability.
\begin{center}
    \begin{mybox} \textbf{RQ4 Summary:}
    \appname demonstrates strong applicability to instruction-tuned LLMs with closed-source datasets, delivering particularly dramatic improvements for initially weaker models.  This expands the method's application scope to broader scenarios where original training datasets are inaccessible, offering a path to enhance LLMs without requiring access to their proprietary training data.
    \end{mybox} 
\end{center}

%% file: Table/main-all.tex
\begin{table}[t]

\caption{Performance of different LLMs using \appname method compared with Standard-SFT on HumanEval(+), MBPP(+) and BigCodeBench, where BCB stands for BigCodeBench.}
\resizebox{\linewidth}{!}{
\begin{tabular}{@{}lcccc@{}}
\toprule
Model & HumanEval(+) & MBPP(+) & \begin{tabular}[c]{@{}c@{}}BCB\\ Full\end{tabular} & \begin{tabular}[c]{@{}c@{}}BCB\\ Hard\end{tabular} \\ \midrule
\multicolumn{5}{l}{\textit{Closed-Source Models}} \\
GPT-3.5-Turbo (Nov 2023) & 76.8 (70.7) & 82.5 (69.7) & 50.6 & 21.6 \\ \midrule
\multicolumn{5}{l}{\textit{Base Model: CodeLlama-Python-7B}} \\
OpenCodeInterpreter-CL & 72.6 (67.7) & 66.4 ({55.4}) & {33.2} & 6.1 \\
AlchemistCoder-CL-7B & {74.4} ({68.3}) & {68.5} (55.1) & 33.1 & {6.7} \\ \hdashline
MagiCoder$\mathcal{S}$-CL & 70.7 (66.5) & 68.4 (56.6) & 39.7 & 12.8 \\
+Standard-SFT & 69.5 (64.0) & 69.3 (58.7) & 39.3 & 13.5 \\
+\appname & \textbf{73.2 (68.9)} & \textbf{71.7 (59.5)} & \textbf{42.2} & \textbf{15.5} \\ \midrule
\multicolumn{5}{l}{\textit{Base Model: DeepseekCoder-6.7B-Base}} \\
WaveCoder-Ultra-DS-6.7B  & 75.0 (69.5) & 74.9 (63.5) & 43.7 & {16.9} \\ 
OpenCodeInterpreter-DS & 76.2 (72.0) & 76.2 ({72.0}) & {44.6} & {16.9} \\
AlchemistCoder-DS-6.7B & {79.9} ({75.6}) & {77.0} (60.2) & 42.5 & 12.2 \\
\hdashline
MagiCoder$\mathcal{S}$-DS & 76.8 (71.3) & 75.7 (64.4) & 47.6 & 12.8 \\
+Standard-SFT & 75.6 (70.7) & 79.1 (66.4) & 46.9 & 10.8 \\
+\appname & \textbf{77.4 (74.3)} & \textbf{79.6 (69.0)} & \textbf{48.2} & \textbf{15.5} \\ \hdashline
SemCoder-S & 79.3 (74.4) & 79.6 (68.5) & 48.5 & 16.9 \\
+Standard-SFT & 79.9 (75.0) & 80.7 (67.2) & 47.1 & 16.2 \\
+\appname & \textbf{83.5 (78.7)} & \textbf{83.1 (70.6)} & \textbf{48.9} & \textbf{20.3} \\ \bottomrule
\end{tabular}%
}

\vspace{-0.1cm}
\label{tab:main-all}
\end{table}

%% file: Table/abl-evalplus.tex
\begin{table}[t]
\caption{Performance ablation of different granularities of differences on HumanEval(+), MBPP(+) and BigCodeBench based on MagiCoder$\mathcal{S}$-DS and SemCoder-S, where BCB stands for BigCodeBench.}
\centering
\resizebox{0.9\linewidth}{!}{
\begin{tabular}{@{}lcccc@{}}
\toprule
Model & HumanEval(+) & MBPP(+) & \begin{tabular}[c]{@{}c@{}}BCB\\ Full\end{tabular} & \begin{tabular}[c]{@{}c@{}}BCB\\ Hard\end{tabular} \\ \midrule
MagiCoderS-DS & 76.8 (71.3) & 75.7 (64.4) & 47.6 & 12.8 \\ \midrule
+\appname (Line Level) & 75.6 (72.0) & 78.8 (68.3) & 47.1 & 14.2 \\
+\appname (Token Level) & 76.2 (72.6) & 79.1 (68.5) & 47.3 & 14.2 \\
+\appname & \textbf{77.4 (74.3)} & \textbf{79.6 (69.0)} & \textbf{48.2} & \textbf{15.5} \\ \midrule
SemCoder-S & 79.3 (74.4) & 79.6 (68.5) & 48.5 & 16.9 \\ \midrule
+\appname (Line Level) & 80.5 (76.8) & \textbf{83.1} (70.1) & \textbf{49.1} & 18.2 \\
+\appname (Token Level) & 81.1 (76.8) & 82.8 (70.1) & 48.3 & 16.9 \\
+\appname & \textbf{83.5 (78.7)} & \textbf{83.1 (70.6)} & 48.9 & \textbf{20.3} \\ \bottomrule
\end{tabular}%
}
\label{tab:abl-evalplus}
\vspace{-0.5cm}
\end{table}

%% file: Table/gen-eval.tex
\begin{table}[t]
\caption{Performance of \appname on other instruction-tuned LLMs with their corresponding datasets on HumanEval(+), MBPP(+) and BigCodeBench, where BCB stands for BigCodeBench.}
\centering
\resizebox{0.9\linewidth}{!}{
\begin{tabular}{@{}lcccc@{}}
\toprule
Model & HumanEval(+) & MBPP(+) & \begin{tabular}[c]{@{}c@{}}BCB\\ Full\end{tabular} & \begin{tabular}[c]{@{}c@{}}BCB\\ Hard\end{tabular} \\ \midrule
\multicolumn{5}{l}{\textit{(Corresponding Dataset OSS-INSTRUCT)}} \\
MagiCoder-DS & 66.5 (60.4) & 75.4 (61.9) & 43.4 & 12.2 \\
+Standard-SFT & 64.6 (58.5) & 79.1 (66.1) & 43.9 & 12.2 \\
+\appname & \textbf{67.1 (62.2)} & \textbf{79.4 (66.4)} & \textbf{46.2} & \textbf{15.5} \\ \midrule
\multicolumn{5}{l}{\textit{(Corresponding Dataset PYX)}} \\
SemCoder & 73.2 (68.9) & 79.9 (65.3) & 43.5 & 16.9 \\
+Standard-SFT & 71.3 (65.2) & 79.9 (66.4) & 43.4 & 14.2 \\
+\appname & \textbf{73.7 (69.5)} & \textbf{81.0 (67.2)} & \textbf{47.9} & \textbf{21.6} \\ \bottomrule
\end{tabular}%
}
\label{tab:gen-eval}
\vspace{-0.5cm}
\end{table}

%% file: Table/rq4-eval.tex
\begin{table}[t]
\caption{Performance of LLMs trained on closed-source instruction-tuning datasets after using \appname on HumanEval(+), MBPP(+) and BigCodeBench, where BCB stands for BigCodeBench.}
\centering
\label{tab:rq4-eval}
\resizebox{0.9\linewidth}{!}{%
\begin{tabular}{@{}lcccc@{}}
\toprule
Model & HumanEval(+) & MBPP(+) & \begin{tabular}[c]{@{}c@{}}BCB\\ Full\end{tabular} & \begin{tabular}[c]{@{}c@{}}BCB\\ Hard\end{tabular} \\ \midrule
\multicolumn{5}{l}{\textit{Base Model: CodeLlama-Python-7B}} \\
CodeLlama-Instruct & 36.0 (31.1) & 56.1 (46.6) & 25.7 & 4.1 \\
+Standard-SFT & 39.0 (34.1) & 61.1 (49.7) & 26.5 & 4.1 \\
+\appname & \textbf{47.0 (43.9)} & \textbf{61.9 (51.3)} & \textbf{29.0} & \textbf{4.7} \\ \midrule
\multicolumn{5}{l}{\textit{Base Model: DeepseekCoder-6.7B-Base}} \\
DeepseekCoder-Instruct & 73.8 (70.7) & 74.9 (65.6) & 43.8 & 15.5 \\
+Standard-SFT & 75.6 (70.1) & 77.8 (66.9) & 42.0 & 13.5 \\
+\appname & \textbf{81.7 (76.2)} & \textbf{78.6 (66.9)} & \textbf{44.3} & \textbf{16.9} \\ \bottomrule
\end{tabular}%
}
\vspace{-0.3cm}
\end{table}

%% file: latex/Dis.tex
\section{Discussion}
\label{Sec:dis}
\begin{figure}[htbp]
    \centering 

    \begin{subfigure}[b]{0.8\linewidth} 
        \centering
        \includegraphics[width=\linewidth]{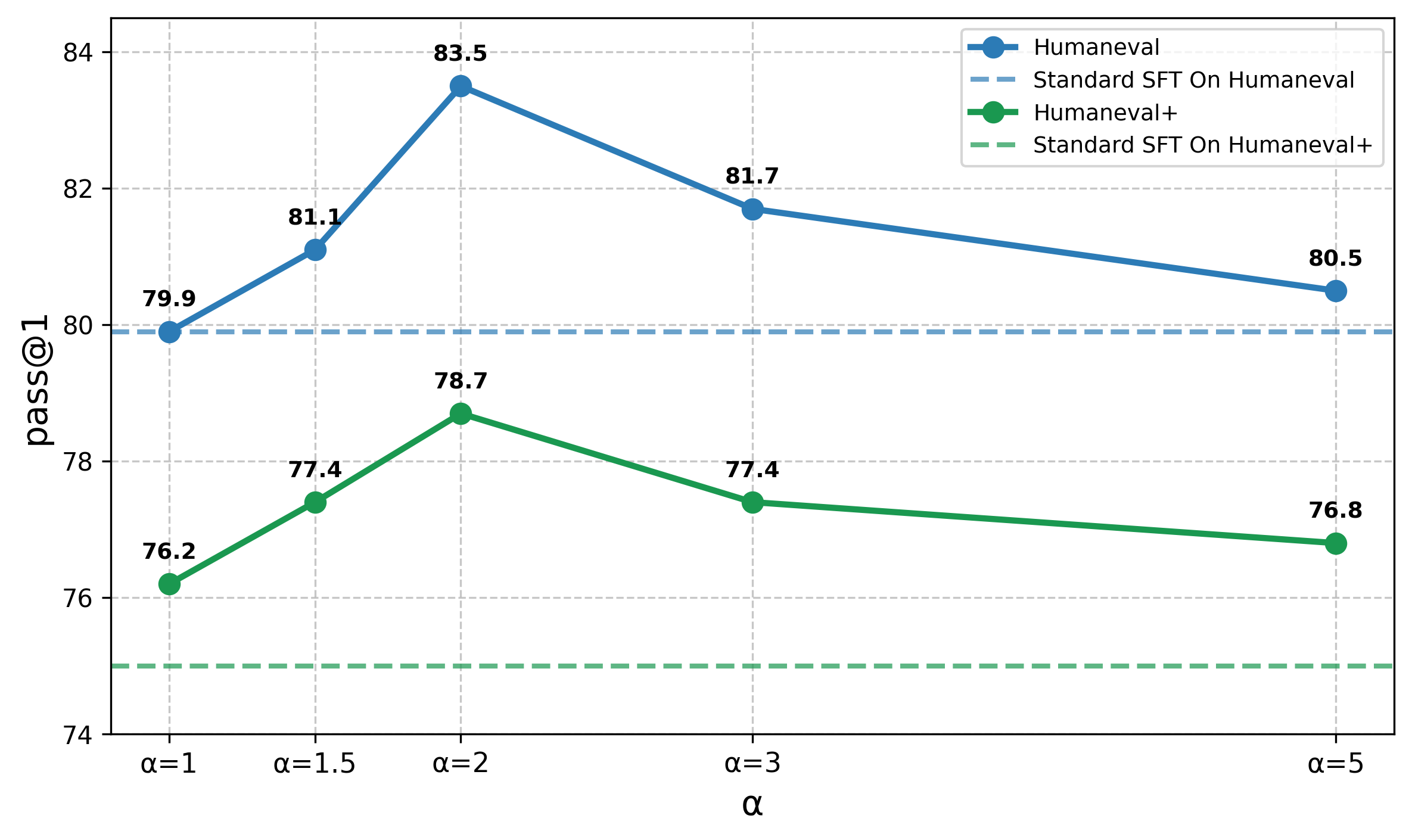} 
        \caption{Performance of different values of $\alpha$ based on SemCoder-S on HumanEval(+).}
        \label{fig:abl-alpha-human}
    \end{subfigure}
    \\
    \begin{subfigure}[b]{0.8\linewidth} 
        \centering
        \includegraphics[width=\linewidth]{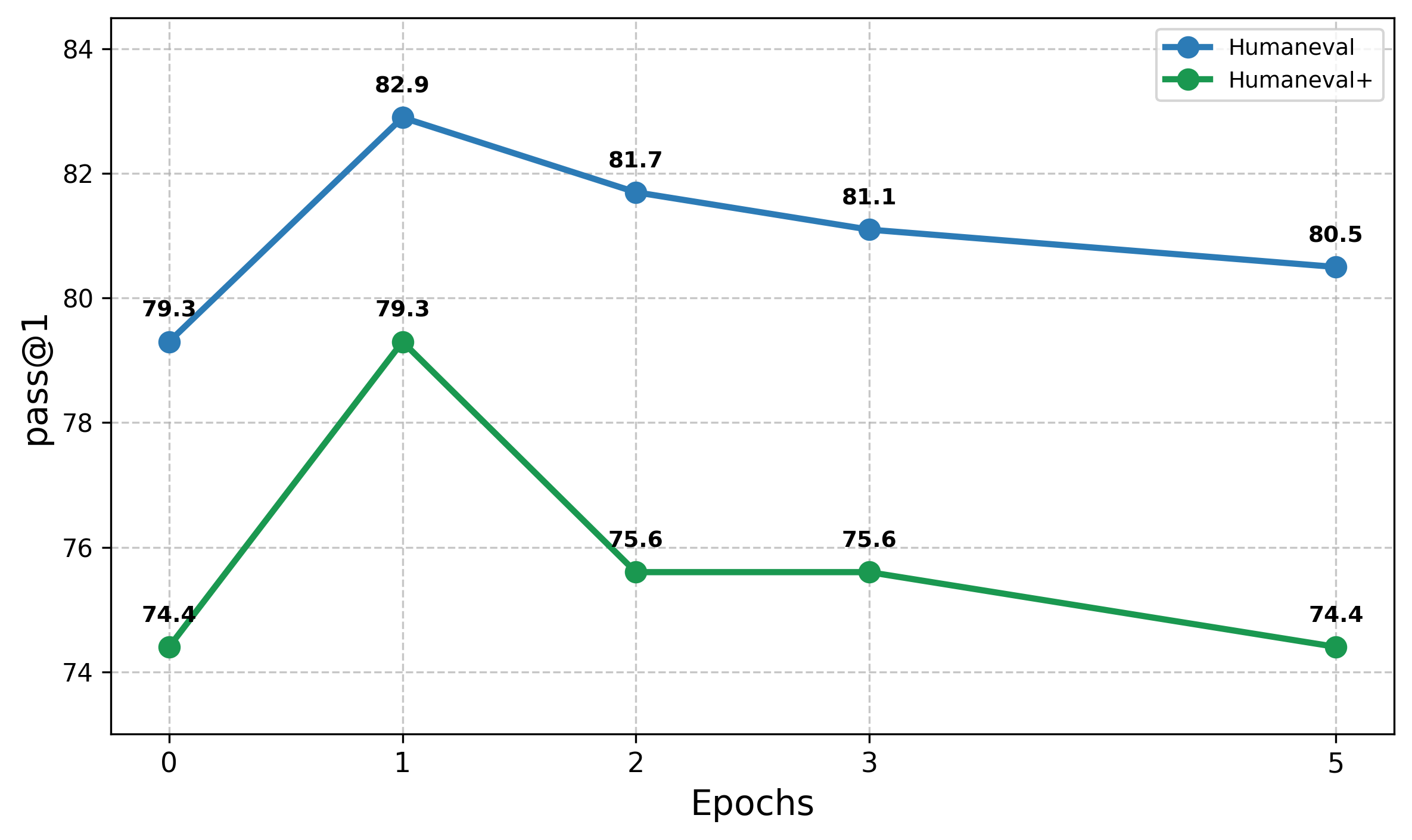} 
        \caption{Performance of different epochs based on SemCoder-S on HumanEval(+).}
        \label{fig:abl-epoch-human}
    \end{subfigure}
    
    \caption{Performance analysis of SemCoder-S on HumanEval(+): varying hyperparameter $\alpha$ (top), and varying training epochs (bottom).}
    \label{fig:combined-performance-analysis}
    \vspace{-0.3cm}
\end{figure}

\textbf{\textit{Impact of hyperparameters}. }
We first analyse the impact of the parameter $\alpha$ in our loss function, which controls the emphasis placed on error-sensitive tokens, by changing $\alpha$. Specifically, we conduct experiments with $\alpha \in [1, 1.5, 2, 3, 5]$ on HumanEval(+) and observe the performance changes of LLMs. Due to the evaluation time constraints, we select SemCoder-S as the representative LLM for this ablation study, as it demonstrates the best overall performance with \appname. Figure~\ref{fig:abl-alpha-human} illustrates the performance trends as $\alpha$ varies. We can observe that $\alpha=2$ provides an optimal balance between emphasizing error-sensitive tokens and maintaining attention to shared tokens. Additionally, it can be observed that in all cases, after \appname, the LLM's performance matches or exceeds that of Standard-SFT, demonstrating the robustness to hyperparameter choices.

Then we investigate the impact of the number of training epochs, which also influences the learning emphasis on \ess. Specifically, we conduct experiments with epochs in $[1, 2, 3, 5]$ using SemCoder-S and observe the performance changes on HumanEval(+). Figure~\ref{fig:abl-epoch-human} illustrates the performance trends as the number of training epochs varies. We find that an excessive number of training epochs can lead to a decline in performance. This might be because the model over-focuses on \ess, neglecting the importance of surrounding tokens.

\input{Table/dis-dpo-all}
\textbf{\textit{Compared to Reinforcement Learning Method}.}
Given the increasing popularity of reinforcement learning (RL) methods for improving code generation~\cite{le2022coderl,wang2022compilable}, we believe it's important to compare our approach with these established techniques. As RL methods in code generation typically aim to align model outputs with desired code solutions by increasing the probability of correct implementations while reducing the likelihood of erroneous ones, they share similarities with our work principle of enhancing LLMs' ability to identify \ess to improve code generation capabilities. Specifically, we compare our approach with the representative DPO method~\cite{rafailov2023direct}, which is widely used and has demonstrated significant advantages in code generation~\cite{yang2024synthesizing,miao2024aligning,gee2024code,zhang2024codedpo,gallego2024refined}.
This method works by training models to directly maximize the likelihood of preferred outputs over non-preferred ones without requiring explicit reward modeling, learning from paired examples of more and less desirable code implementations.

To ensure a fair comparison, we select the same LLMs and use identical experimental settings as stated in RQ1 for DPO training. The training dataset remains consistent across both DPO and \appname, and we evaluate and compare the performance of DPO and \appname on HumanEval(+), MBPP(+), and BigCodeBench. Table ~\ref{tab:dis-dpo-eval} shows the performance of LLMs trained with different methods. Overall, LLMs with \appname consistently outperform those trained with DPO. We can observe that \appname outperforms DPO by a relative average of 4.2\%, across three selected benchmarks. This advantage stems from fundamental methodological differences: while DPO relies on coarse-grained preference signals that cannot precisely target \ess, \appname specifically maintains learning across all tokens while strategically emphasizing \ess within code implementations. This approach ensures the LLM retains general coding knowledge while becoming more attentive to critical details that often determine functional correctness.

Additionally, we note that DPO's ability to differentially reward correct implementations and penalize incorrect ones could be leveraged to enhance the learning of error-sensitive segments. Specifically, a tailored reward function could be designed to strengthen the model's focus on these critical segments, potentially combining the strengths of both approaches. We leave this promising direction for future exploration.

\textbf{\textit{Analysis of Overfitting Dynamics}.}
\begin{figure}[t]
    \centering 

    \begin{subfigure}[b]{0.32\linewidth} 
        \centering
        \includegraphics[width=\linewidth]{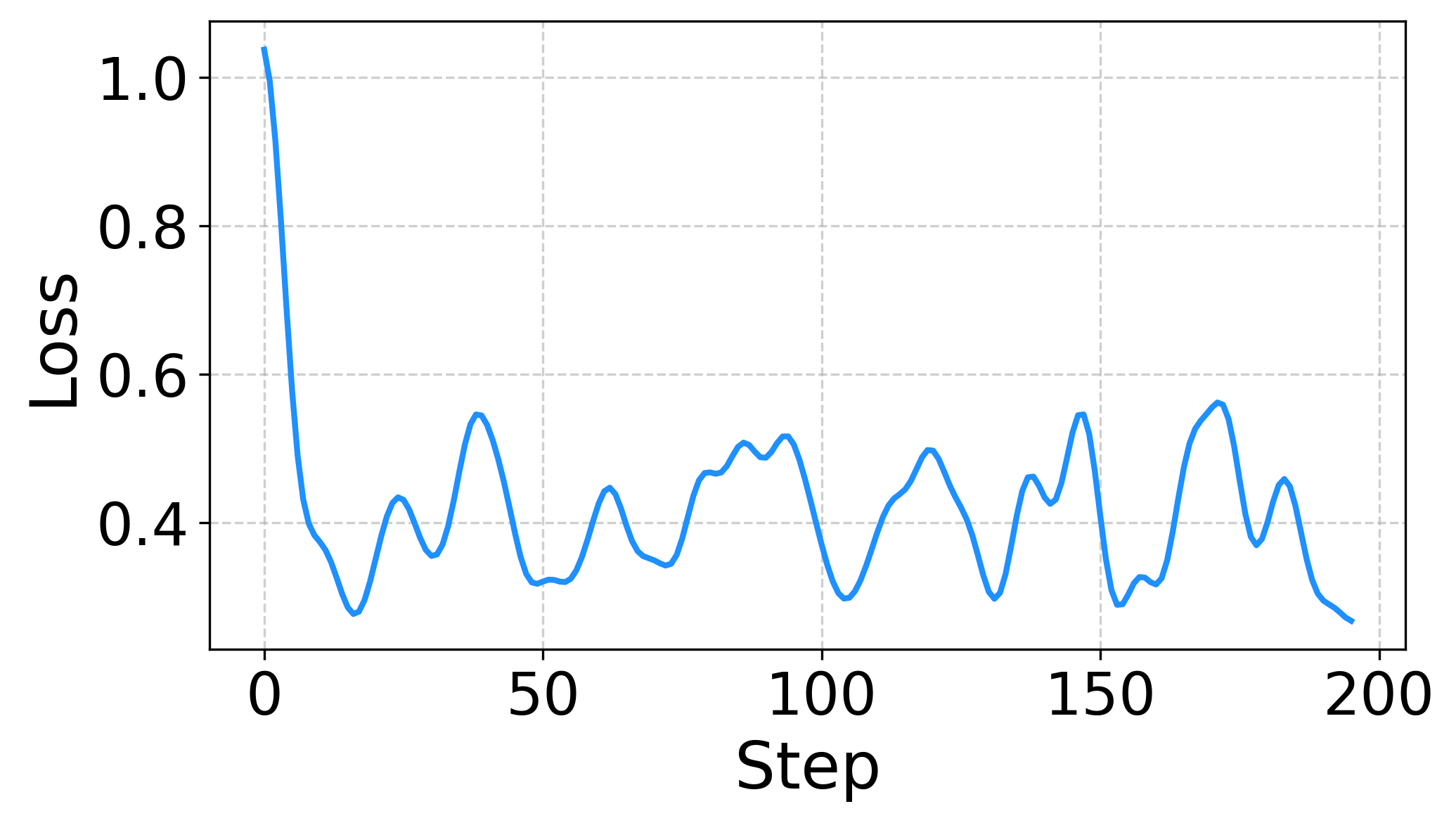} 
        \caption{}
        \label{fig:trainloss}
    \end{subfigure}
    \hfill 
    \begin{subfigure}[b]{0.32\linewidth} 
        \centering
        \includegraphics[width=\linewidth]{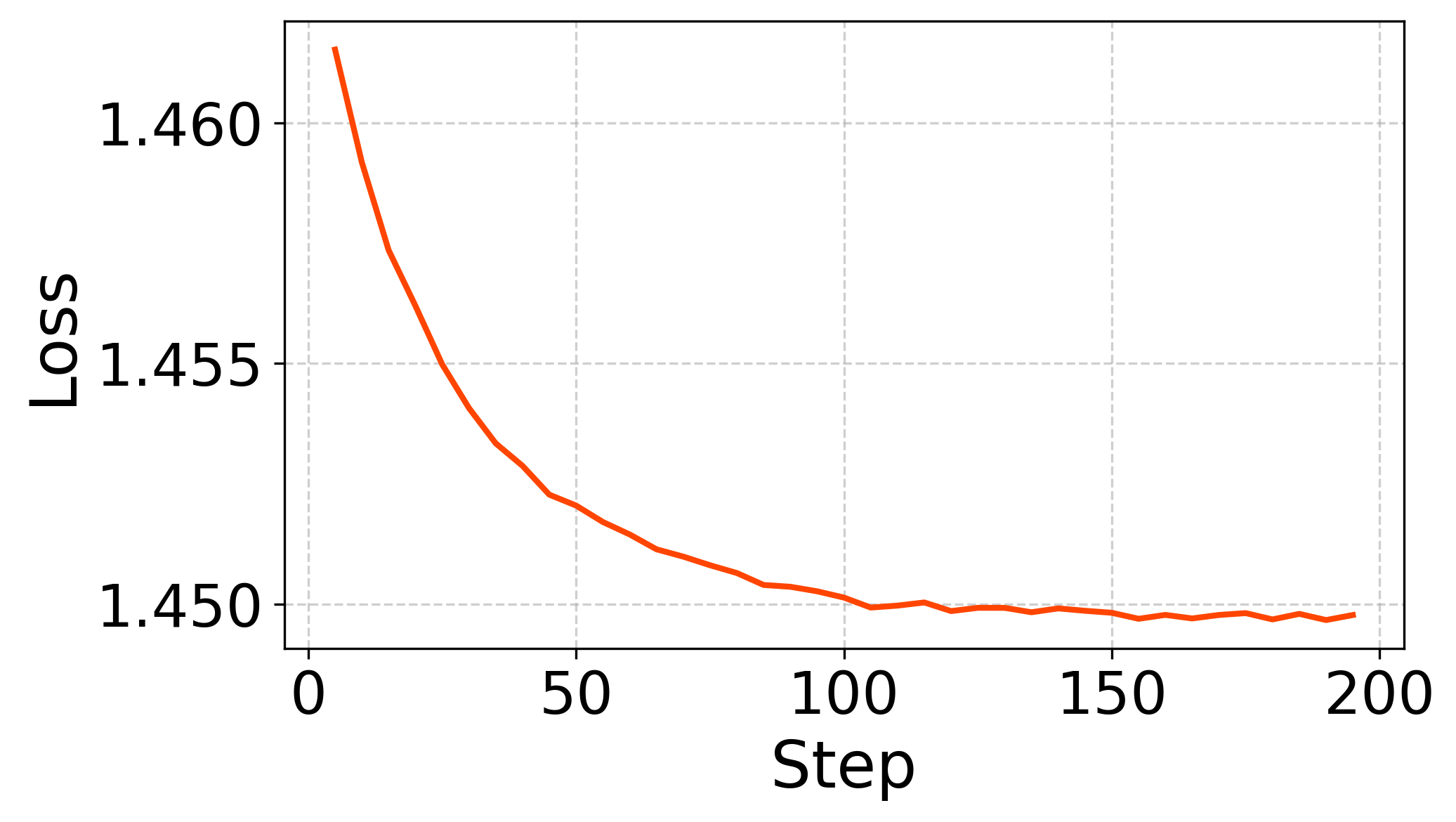} 
        \caption{}
        \label{fig:valloss}
    \end{subfigure}
    \hfill 
    \begin{subfigure}[b]{0.32\linewidth} 
        \centering
        \includegraphics[width=\linewidth]{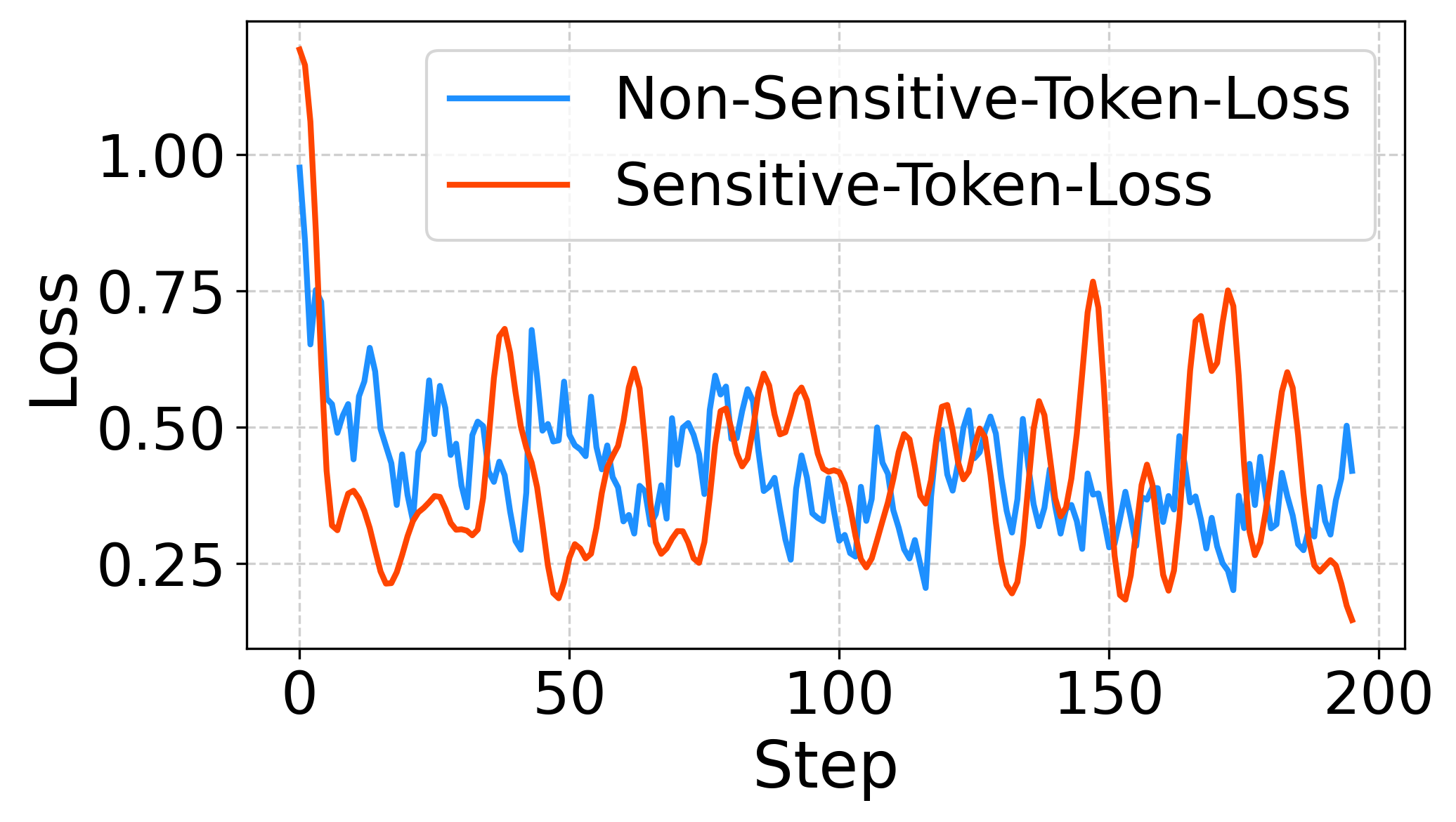} 
        \caption{}
        \label{fig:tokenloss}
    \end{subfigure}
    
    \label{fig:allloss}
    \caption{Learning curves for SemCoder-S with \appname on Evol-Instruct. (a) training loss, (b) validation loss, and (c) training loss categorized by token type.}
\end{figure}
To investigate the possibility of potential overfitting, we analyze the learning dynamics. Specifically, we randomly partition Evol-Instruct into a 9:1 training-validation split and reduce the batch size to obtain a finer-grained view of the loss curves. Figure~\ref{fig:trainloss} and Figure~\ref{fig:valloss} illustrate the learning curves for the SemCoder-S model. Notably, the validation loss for \appname shows a consistent decrease, suggesting \appname mitigates the overfitting issue. We hypothesise that by assigning higher weights to more challenging, error-sensitive tokens, \appname guides the model to focus on harder-to-learn patterns, thus mitigating overfitting. Furthermore, to ensure comprehensive learning across the entire code structure, we assign a non-zero base weight ($\alpha-1$) to non-sensitive tokens. Figure~\ref{fig:tokenloss} confirms that the loss for both sensitive and non-sensitive tokens decreases during training.

\textbf{\textit{Comparison with Mutation-Based Method}.}
\input{Table/dis-muta}
Traditional mutation-based approaches provide a natural baseline for our method, as they are also designed to introduce errors into code. We replace the LLM-based method with the rule-based universalmutator~\cite{deb2024syntax} to generate faulty versions of the code in Evol-Instruct. It is a state-of-the-art, language-agnostic mutation tool. We train SemCoder-S using these mutated samples, keeping all hyperparameters consistent. As shown in Table~\ref{tab:muta}, \appname outperforms the mutation-based approach. We attribute this to the fact that LLMs, trained on vast code corpora, generate a more diverse and realistic spectrum of plausible errors than a fixed set of mutation rules, leading to more effective learning.

\input{Table/dis-overfit}
\textbf{\textit{Potential Overfitting of Standard-SFT}.}
We observe that Standard-SFT occasionally lead to performance degradation compared to the base model. For instance, as detailed in Table~\ref{tab:gen-eval}, SemCoder exhibit a decline in performance across multiple benchmarks after undergoing Standard-SFT. 
We hypothesize that this can be attributable to potential overfitting. The model is trained on PYX whose data distribution is already familiar from its initial fine-tuning phase.
To validate this, we fine-tune models on datasets they have not previously been exposed to. Specifically, we fine-tune MagiCoder-DS on PYX and Evol-Instruct, and SemCoder on OSS-Instruct and Evol-Instruct. As shown in Table~\ref{tab:overfit}, the models achieve more gains when trained on unseen datasets. For example, SemCoder, when fine-tuned on either OSS-Instruct or Evol-Instruct, outperforms the version trained on its original dataset, PYX.

%% file: Table/dis-dpo-all.tex
\begin{table}[t]
\caption{Performance of different LLMs using \appname method compared with DPO on HumanEval(+), MBPP(+) and BigCodeBench, where BCB stands for BigCodeBench.}
\label{tab:dis-dpo-eval}
\centering
\resizebox{0.9\linewidth}{!}{%
\begin{tabular}{@{}lcccc@{}}
\toprule
Model & HumanEval(+) & MBPP(+) & \begin{tabular}[c]{@{}c@{}}BCB\\ Full\end{tabular} & \begin{tabular}[c]{@{}c@{}}BCB\\ Hard\end{tabular} \\ \midrule
\multicolumn{5}{l}{\textit{Base Model: CodeLlama-Python-7B}} \\
MagiCoder$\mathcal{S}$-CL & 70.7 (66.5) & 68.4 (56.6) & 39.7 & 12.8 \\
+DPO & 66.5 (61.6) & 68.8 (58.7) & 39.8 & 14.2 \\
+\appname & \textbf{73.2 (68.9)} & \textbf{71.7 (59.5)} & \textbf{42.2} & \textbf{15.5} \\ \midrule
\multicolumn{5}{l}{\textit{Base Model: DeepseekCoder-6.7B-Base}} \\
MagiCoder$\mathcal{S}$-DS & 76.8 (71.3) & 75.7 (64.4) & 47.6 & 12.8 \\
+DPO & 76.2 (71.9) & 79.1 (68.3) & 47.8 & 13.5 \\
+\appname & \textbf{77.4 (74.3)} & \textbf{79.6 (69.0)} & \textbf{48.2} & \textbf{15.5} \\ \hdashline
SemCoder-S & 79.3 (74.4) & 79.6 (68.5) & 48.5 & 16.9 \\
+DPO & 81.7 (76.2) & 81.0 (67.7) & 47.9 & 16.2 \\
+\appname & \textbf{82.9 (79.3)} & \textbf{83.1 (70.6)} & \textbf{48.9} & \textbf{20.3} \\ \bottomrule
\end{tabular}%
}
\vspace{-0.3cm}
\end{table}

%% file: Table/dis-muta.tex
\begin{table}[]
\caption{Performance comparison between \appname and the traditional mutation-based method on HumanEval(+), MBPP(+) and
BigCodeBench, where BCB stands for BigCodeBench.}
\label{tab:muta}
\resizebox{\linewidth}{!}{%
\begin{tabular}{@{}lccccc@{}}
\toprule
Method & Humaneval (+) & MBPP(+) & \begin{tabular}[c]{@{}c@{}}BCB\\ Full\end{tabular} & \begin{tabular}[c]{@{}c@{}}BCB\\ Hard\end{tabular} & Avg \\ \midrule
Original & 79.3 (74.4) & 79.6 (68.5) & 48.5 & 16.9 & 61.2 \\
Mutation-Based & 81.1 (77.4) & 80.9 (67.7) & \textbf{49.2} & 17.5 & 62.3 \\
LLM-Based (FGIT) & \textbf{83.5 (78.7)} & \textbf{83.1 (70.6)} & 48.9 & \textbf{20.3} & \textbf{64.2} \\ \bottomrule
\end{tabular}%
}
\end{table}

%% file: Table/dis-overfit.tex
\begin{table}[t]
\caption{Performance Impact of Standard-SFT on seen and unseen datasets on HumanEval(+), MBPP(+) and BigCodeBench, where BCB stands for BigCodeBench.}
\label{tab:overfit}
\resizebox{\linewidth}{!}{%
\begin{tabular}{@{}llccccc@{}}
\toprule
Model & Dataset & Humaneval (+) & MBPP (+) & \begin{tabular}[c]{@{}c@{}}BCB\\ Full\end{tabular} & \begin{tabular}[c]{@{}c@{}}BCB\\ Hard\end{tabular} & Avg \\ \midrule
\multirow{4}{*}{MagiCoder-DS} & None & 66.5 (60.4) & 75.4 (61.9) & 43.4 & 12.2 & 53.3 \\
 & OSS-Instruct (Original) & 64.6 (58.5) & \textbf{79.1 (66.1)} & 43.9 & 12.2 & 54.1 \\
 & PYX (New) & \textbf{66.5 (60.9)} & 77.5 (65.8) & 45.7 & \textbf{15.5} & \textbf{55.3} \\
 & Evol-Instruct (New) & 64.6 (60.9) & 77.2 (65.6) & \textbf{47.3} & \textbf{15.5} & 55.2 \\
\multirow{4}{*}{SemCoder} & None & 73.2 (68.9) & 79.9 (65.3) & 43.5 & 16.9 & 58 \\
 & PYX (Original) & 71.3 (65.2) & \textbf{79.9 (66.5)} & 43.4 & 14.2 & 56.8 \\
 & OSS-Instruct (New) & \textbf{73.2 (69.5)} & 78.0 (64.3) & 46.4 & \textbf{18.9} & \textbf{58.4} \\
 & Evol-Instruct (New) & 71.3 (67.6) & 78.6 (65.6) & \textbf{47.6} & \textbf{19.6} & \textbf{58.4} \\ \bottomrule
\end{tabular}%
}
\end{table}

%% file: latex/Threat.tex
\section{Threats To Validity}
\label{sec:6}
\textit{Threats to external validity}  relate to the generalizability of our approach. While we evaluate our approach on multiple instruction-tuned models, there may be concerns about generalization to other LLMs. However, this threat is mitigated by our diverse selection of models with different series. Furthermore, the cross-dataset experiments (RQ3) and closed-source dataset experiments (RQ4) demonstrate robust generalization capabilities across different settings. In addition, due to computational resource constraints, our experiments primarily focus on 7B parameter LLMs rather than larger LLMs. In future work, we plan to explore a broader range of model series to further validate our approach's generalizability.

\textit{Threats to internal validity} involve the impact of the quality of incorrect code and choices of hyperparameters. The effectiveness of our approach depends on the quality of incorrect code variants, the weighting factor $\alpha$ and training epochs. To mitigate threats related to code quality, we prompt a strong teacher model to generate plausible incorrect variants. Subsequently, we manually and quantitatively analyze the portion of generated data which is incorrect yet similar to the correct solutions. While a small portion of noise data remains present, we argue these instances may actually enhance model robustness by preventing overfitting to specific \ess~\cite{devries2017improved,xie2020unsupervised}.  For hyperparameter-related threats, we conduct an extensive sensitivity analysis as shown in Figure~\ref{fig:combined-performance-analysis}. In future work, we intend to investigate the use of stronger teacher models, such as GPT-4-Turbo, to generate similar incorrect code and examine their impact.

%% file: latex/Related.tex
\section{Related Work}
\label{sec:2}
\subsection{LLMs for Code Generation}
LLMs are increasingly being leveraged to automate various tasks in software engineering ~\cite{ahmad2023towards,pan2024automating,fan2025exploring,nijkampcodegen,fan2025sek}. Among these, code generation has emerged as a particularly prominent area of research and application~\cite{li2022competition,nijkampcodegen,friedincoder,li2023starcoder}. As a momentous milestone, Codex~\cite{chen2021evaluating} boasting a 12-billion-parameter model demonstrates the extraordinary capability to tackle up to 72\% of Python programming problems. After that, a new wave of code generation models, such as AlphaCode~\cite{li2022competition}, CodeGen~\cite{nijkampcodegen}, InCoder~\cite{friedincoder} and StarCoder~\cite{li2023starcoder} are proposed and have shown promising results in the code generation task. Building upon these foundations, more code-focused LLMs emerged, such as Magicoder~\cite{wei2024magicoder}, SemCoder~\cite{dingsemcoder}, WaveCoder~\cite{yu2024wavecoder} and WizardCoder~\cite{luowizardcoder}. These specialized LLMs are typically based on general LLMs in solving domain-specific coding tasks through instruction tuning. 

\subsection{Fine-tuning on Code LLM}
Fine-tuning pre-trained language models has emerged as a dominant paradigm for optimizing performance in code generation. Instruction tuning~\cite{ouyang2022training,achiam2023gpt}, as a form of supervised fine-tuning, aims to align LLMs with instruction through high-quality instruction corpora. For instance, Magicoder~\cite{wei2024magicoder} introduce OSS-Instruct, a dataset generated by a teacher LLM drawing inspiration from open-source code snippets, which effectively enhances code generation capabilities.
Similarly, SemCoder~\cite{dingsemcoder} propose PYX, a dataset created by a teacher LLM simulating human debugging processes. By incorporating data that simulates execution reasoning and captures code execution nuances, LLMs finetuned with PYX understand and articulate the execution process step-by-step, enhancing their reasoning capabilities. 
To address limitations in preventing untruthful and unexpected outputs, researchers explore reinforcement learning~\cite{ouyang2022training}. To address limitations in undesired outputs, researchers have explored reinforcement learning approaches using preference optimization~\cite{yang2024synthesizing,shojaeeexecution}, like DPO~\cite{rafailov2023direct}, to refine outputs. However, these often treat tokens uniformly in their loss calculations,  making it difficult for models to distinguish semantically correct implementations from syntactically similar but incorrect ones. In this paper, we aim to address this challenge in code LLMs.

We find recent work Focused-DPO~\cite{zhang2025focused} enhances code generation capability by concentrating preference optimization on error-prone points through an improved DPO~\cite{rafailov2023direct} methodology. Our method is complementary to this approach - while Focused-DPO operates during post-training reinforcement learning stages, our approach operates during the supervised fine-tuning stage. 
However, as Focused-DPO is under peer review and its implementation is not yet publicly available, we are unable to experimentally validate the potential synergies between our approaches in this work.

%% file: latex/Conclusion.tex
\section{Conclusion}
\label{sec:7}
In this paper, we introduce Fault-Guided Fine-Tuning (\appname), a novel fine-tuning technique that enhances code generation capabilities in instruction-tuned LLMs by refining their ability to distinguish between correct implementations and subtly incorrect variants. 
Through extensive experiments across seven LLMs and three widely-used benchmarks, we demonstrate that our method achieves an average relative improvement of 6.9\% on pass@1, with certain enhanced 6.7B LLMs even outperforming GPT-3.5-Turbo on selected benchmarks. The technique also exhibits strong generalization capabilities across diverse instruction-tuned LLMs and maintains effectiveness even when applied to LLMs with closed-source instruction datasets.

\section*{Acknowledgment}
This research/project was supported by the National Natural Science Foundation of China (No.62202420 and No. 62302430), Zhejiang Provincial Natural Science Foundation of China (No.LZ25F020003 and No.LQ24F020017).

\section*{Data Availability}
Our code is available: https://github.com/ZJU-CTAG/FGIT.